\documentclass[aip,apl,amsmath,amssymb,reprint,unsortedaddress,superscriptaddress]{revtex4-1}
\setcitestyle{super}

\usepackage{graphicx}% Include figure files
\usepackage{dcolumn}% Align table columns on decimal point
\usepackage{bm}% bold mat
\usepackage{textgreek}
\usepackage{color,soul}% mark text by wrapping it with \hl{}
\usepackage{comment}
\usepackage{multirow}%use for multirows in tables
\usepackage[normalem]{ulem} % Allow for strikout texts using \sout{}

\makeatletter
\let\@fnsymbol\@fnsymbol@latex
\@booleanfalse\altaffilletter@sw
\makeatother

\begin{document}
	
\preprint{AIP/123-QED}
	
\title[\textsc{Current-induced switching of YIG/Pt bilayers with in-plane magnetization due to Oersted fields}]{Current-induced switching of YIG/Pt bilayers with in-plane magnetization due to Oersted fields}% Force line breaks with \\
	
\author{Johannes Mendil}\email{johannes.mendil@mat.ethz.ch}
\affiliation{Department of Materials, ETH Zurich, 8093 Zurich, Switzerland}
	
\author{Morgan Trassin}
\affiliation{Department of Materials, ETH Zurich, 8093 Zurich, Switzerland}
	
\author{Quingquing Bu}
\affiliation{Department of Materials, ETH Zurich, 8093 Zurich, Switzerland}
	
\author{Manfred Fiebig}
\affiliation{Department of Materials, ETH Zurich, 8093 Zurich, Switzerland}
	
\author{Pietro Gambardella}
\affiliation{Department of Materials, ETH Zurich, 8093 Zurich, Switzerland}
	
\date{\today}
	
\begin{abstract}
We report on the switching of the in-plane magnetization of thin yttrium iron garnet (YIG)/Pt bilayers induced by an electrical current. The switching is either field-induced and assisted by a dc current, or current-induced and assisted by a static magnetic field. The reversal of the magnetization occurs at a current density as low as $10^5$~A/cm$^{2}$ and magnetic fields of $\sim 40$~$\mu$T, two orders of magnitude
smaller than in ferromagnetic metals, consistently with the weak uniaxial anisotropy of the YIG layers. We use the transverse component of the spin Hall magnetoresistance to sense the magnetic orientation of YIG while sweeping the current. Our measurements and simulations reveal that the current-induced effective field responsible for switching is due to the Oersted field generated by the current flowing in the Pt layer rather than by spin-orbit torques, and that the switching efficiency is influenced by pinning of the magnetic domains.

\end{abstract}
\maketitle

%%%%%%%%%%%%%%%%%%%%%%%%%%%%% Introduction %%%%%%%%%%%%%%%%%%%%%%%%%%%%%%%%%%%%
The possibility of manipulating the magnetization of planar structures using electrical currents opens exciting perspectives in spintronics. Electrical currents can affect the magnetization of thin films through the Oersted magnetic field,\cite{Morecroft2007,Yuan2009,Vogel2011,Nam2011,Fuhrer2011} spin transfer torques,\cite{Brataas2012} and spin-orbit torques.
\cite{Garello2013} Previous work has focused on magnetization switching and domain wall dynamics induced by spin-orbit torques in metallic ferromagnets adjacent to a heavy metal layer.\cite{manchon2018review,Miron2010,Miron2011b,Avci2012,Liu2012,Emori2013,Ryu2013,Baumgartner2017} Recently, investigations extended towards insulating ferrimagnetic garnets, which, owing to the low magnetic damping, are particularly appealing for generating and transmitting spin waves\cite{Uchida2010b,Cornelissen2015,Goennenwein2015,Evelt2016} as well as for magnetization switching.\cite{Avci2017,Avci2017b,Shao2018} The most prominent exponent of this material class is yttrium iron garnet (YIG). Extensive work on the interplay of current-induced effects and magnetization dynamics in YIG/Pt bilayers demonstrated efficient spin-wave excitations,\cite{Schreier2015,Sklenar2015,Lauer2016b,Collet2016,Demidov2016} spin-wave amplification,\cite{Padron2011,Lauer2016a} and the control of magnetization damping.\cite{Wang2011b} So far, however, no attempt at current-induced magnetization switching of YIG has been reported. Two plausible reasons for the scarcity of results in this area are the extreme sensitivity of YIG to magnetic fields, which makes it difficult to control the intermediate magnetization states, as well as to the need to utilize YIG films with uniaxial in-plane anisotropy, which is required to achieve binary switching. Indeed, the electrical switching of garnet insulators has been reported only for thin films with relatively large perpendicular anisotropy, such as thulium iron garnet layers in combination with either Pt or W.\cite{Avci2017,Avci2017b,Shao2018}

In this paper, we investigate the reciprocal effects of current and magnetic field on the switching of YIG/Pt bilayers with in-plane magnetic anisotropy. We demonstrate field-induced switching assisted by a dc current as well as current-induced switching assisted by a static magnetic field at extremely low current density ($\sim 10^5$~A/cm$^{2}$) and bias fields ($40-60$~$\mu$T). We further show that the magnetization reversal can be sensed electrically by measuring the transverse component of the spin Hall magnetoresistance (SMR)\cite{Nakayama2013,Chen2013b,Vlietstra2013b} and adding an ac modulation to the dc current inducing the switching. Current and thickness dependent measurements reveal that the effective switching field is consistent with the Oersted field generated by the current flowing in the Pt layer. No significant effect of spin-orbit torques was detected in the current range from 1 to 8 $\times10^5$ A/cm$^2$ investigated in this work. Our results are relevant for the operation of YIG-based spintronic devices at very low current density in the thin film regime.

\begin{figure}[hbtp!]
\includegraphics[width = 8.5 cm]{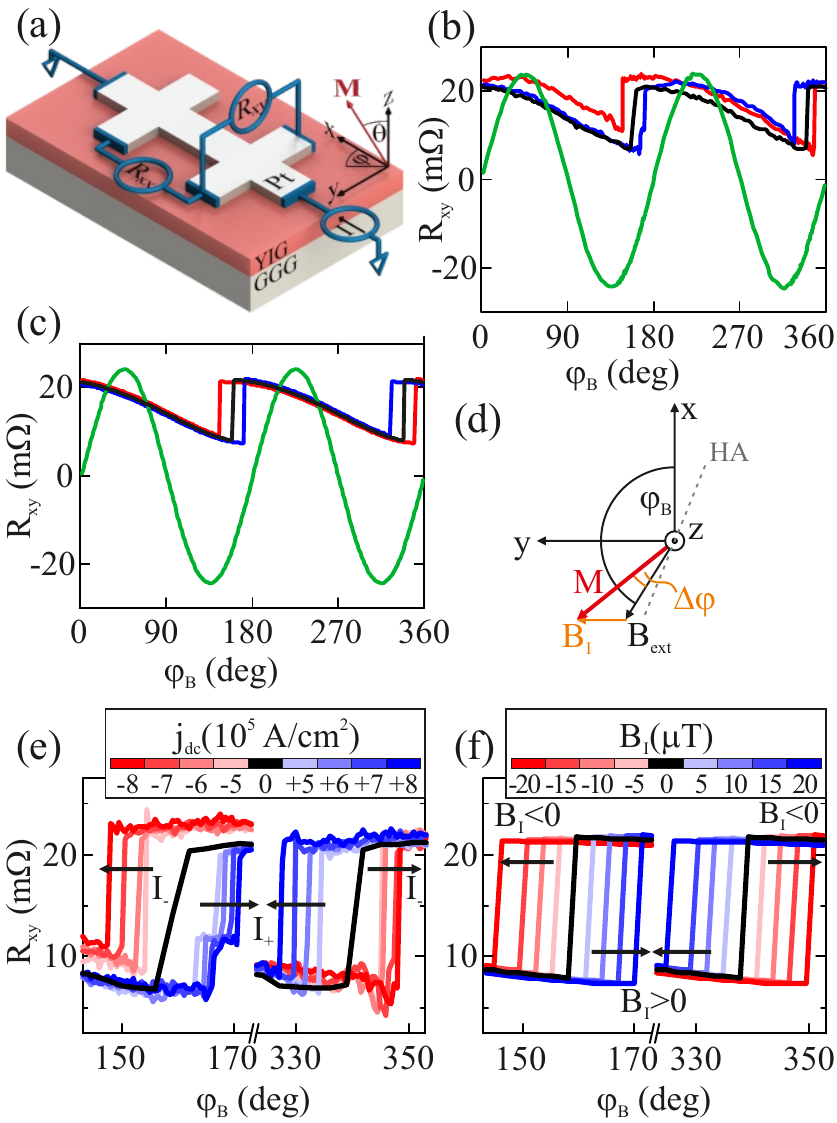}
\caption{\label{fig:current_shift} (a) Schematics of the YIG/Pt Hall bar with the coordinate system. (b) $R_{\rm xy}$ of YIG(6~nm)/Pt(3~nm) measured as a function of $\varphi_{\text{B}}$ at $B_{\text{ext}}= 7$~mT (green line) and 60~$\mu$T (black line). The red an blue lines are measured at $B_{\text{ext}}= 60$~$\mu$T in the presence of a dc offset of 8 and -8$\times10^5$~A/cm$^2$, respectively. (c) Macrospin simulations of the data shown in (b). (d) Diagram showing the combined effect of $B_{\text{I}}$ and $B_{\text{ext}}$ on magnetization switching in proximity of the hard axis (HA). (e) Detail of the shift of $R_{\rm xy}$ as a function of dc offset and (f) macrospin simulations.}
\end{figure}

%%%%%%%%%%%%%%%%%%%%%%%%%%%%% Exp %%%%%%%%%%%%%%%%%%%%%%%%%%%%%%%%%%%%
YIG layers with thickness between 6 and 7~nm were grown epitaxially by pulsed laser deposition on (111)-oriented gadolinium gallium garnet substrates, followed by in-situ magnetron sputtering of a 3~nm thick polycrystalline Pt film with a sheet resistance of $160$~$\Omega$. For electrical measurements, the samples were patterned into Hall bars using optical lithography followed by Ar-ion milling [Fig.~\ref{fig:current_shift}~(a)]. The current line is 50~$\mu$m wide and is oriented along the $[1\bar{1}0]$ crystal direction of the substrate. The separation between two consecutive Hall arms is 500~$\mu$m.
The YIG layers have in-plane magnetization with saturation value $M_s = (1.0\pm 0.2)\times 10^5$~A/m, which is smaller by about 30\% compared to the $M_s$ of bulk YIG. This reduced $M_s$, typical for very thin YIG, is assigned to the diffusion of Gd atoms from the substrate into YIG.\cite{Mendil2019} In addition to the shape anisotropy, the layers have a rather strong easy plane anisotropy, corresponding to an effective isotropic anisotropy field of about 75~mT, and a weaker in-plane uniaxial anisotropy, corresponding to an in-plane anisotropy field $B_{\text{K}} \approx 40-50$~$\mu$T,
%which defines the equilibrium orientation of the magnetization and
which is not correlated to a specific crystal direction. The origin of the uniaxial in-plane anisotropy in the epitaxial YIG(111) layers is attributed to local strain variations introduced during the microfabrication process. A detailed structural and magnetic characterization of our samples is reported in Ref.~\onlinecite{Mendil2019}.

To sense the magnetic orientation and current-induced effective fields, we performed harmonic Hall voltage measurements,\cite{Garello2013,Avci2014b} whereby an ac current with a frequency of 10~Hz and current density $j=10^5$ A/cm$^2$ is sent through the Hall bar while the transverse resistance is acquired and decomposed into its harmonic components. To derive the orientation of the in-plane magnetization, it is sufficient to consider the first harmonic Hall resistance $R_{\text{xy}}$ as a function of the direction of the external magnetic field $B_{\text{ext}}$. The azimuthal angles of $B_{\text{ext}}$ and magnetization are $\varphi_{\text{B}}$ and $\varphi$, respectively, defined with respect to the current direction. The corresponding polar angles are $\theta_{\text{B}}$ and $\theta$ [see Fig.~\ref{fig:current_shift}~(a)]. $B_{\text{ext}}$ is measured by a calibrated Hall sensor placed next to the sample, without correction for the earth's magnetic field. All experiments are performed at room temperature.

Figure~\ref{fig:current_shift}~(b) shows $R_{\text{xy}}$ of YIG(6~nm)/Pt(3~nm) measured as a function of $\varphi_{\text{B}}$ for $B_{\text{ext}}=7$~mT (green curve) and 60~$\mu$T (black curve). As $\theta=\theta_{\text{B}}=\pi/2$, the Hall resistance is determined by the planar Hall-like contribution from the SMR\cite{Nakayama2013,Chen2013b}
\begin{align}
\label{eq:Rphe}
R_{\text{xy}}=R_{\perp}\sin(2\varphi),
\end{align}
where $R_{\perp}$ denotes the transverse SMR coefficient. If the magnetization is saturated parallel to the field, we have that $\varphi=\varphi_{\rm B}$ and $R_{\text{xy}}=R_{\perp}\sin(2\varphi_{\text{B}})$, in agreement with the measurement performed at $B_{\text{ext}}=7$~mT. Conversely, for $B_{\text{ext}}=60$~$\mu$T, that is, comparable or smaller than $B_{\rm K}$, we observe significant deviations from the saturated behavior. These deviations consist in a reduction of the signal amplitude, due to $\varphi \neq \varphi_{\rm B}$, and two abrupt jumps separated by 180$^{\circ}$. We attribute these jumps to the sudden switch of the magnetization from the positive to the negative direction (relative to the easy axis) as $B_{\text{ext}}$ crosses the hard axis, consistently with the uniaxial in-plane anisotropy of our films.

In order to support this hypothesis and quantify $B_{\rm K}$, we performed macrospin simulations based on the magnetic energy functional
\begin{align}
E=-\mathbf{M}\cdot\mathbf{B}_{\text{ext}}+M_{\rm s}B_{\rm K}\sin^2(\varphi-\varphi_{\rm EA})-\mathbf{M}\cdot\mathbf{B}_{\text{I}},
\label{eq:E}
\end{align}
where the first two terms on the right hand side correspond to the Zeeman energy and uniaxial in-plane anisotropy energy, respectively, and the last term represents the interaction between the magnetization $M$ and the current-induced magnetic field $B_{\text{I}}$, which we will discuss later on. Minimization of $E$ for a given set of $\varphi_{\rm B}$ at constant $B_{\text{ext}}$ and $B_{\rm I}=0$ yields $B_{\rm K}$ and a set of values $\varphi$, which we use to simulate $R_{\rm xy}$ using Eq.~(\ref{eq:Rphe}). The best fit between simulations and data is achieved for $B_{\rm K}=40$~$\mu$T and an easy axis $\varphi_{\rm EA}=63^{\circ}$. The $R_{\rm xy}$ curves calculated using these parameters are shown in Fig.~\ref{fig:current_shift}~(c) for different values of $B_{\text{ext}}$. The simulations reproduce fairly well the main features of the Hall resistance measurements, namely the lineshape, the amplitude and position of the jumps, and their separation by $180^{\circ}$. We thus conclude that the macrospin model is appropriate to describe the behavior of the magnetization, at least in the Hall cross region probed by $R_{\rm xy}$.

Since $B_{\text{ext}}$ and $B_{\text{K}}$ are in the range of tens of $\mu$T, we expect that any additional current-induced field $B_{\text{I}}$ should have a pronounced impact on the orientation of the magnetization, even for very small current densities. To prove this point, we added a dc offset to the ac current and measured $R_{\rm xy}$ at low field as a function of $\varphi_{\rm B}$. For a dc offset of $8\times 10^{5}$~A/cm$^2$, we observe that the angle $\varphi_{\rm B}$ at which the magnetization switches shifts by an amount $\Delta\varphi$. The sign of $\Delta\varphi$ depends on the polarity  of the dc current, as shown by the red and blue curves in Fig.~\ref{fig:current_shift}~(b). Such a shift is attributed to the action of a dc field $B_{\rm I}$, which assists $B_{\text{ext}}$ such as to favor or hinder the switching of the magnetization in proximity of the hard axis [Fig.~\ref{fig:current_shift}~(d)]. Accordingly, in the first hemicycle ($0^{\circ}\leq\varphi_{\text{B}}<180^{\circ}$), a negative (positive) current shifts the magnetization reversal towards smaller (larger) $\varphi_{\text{B}}$, whereas, in the second hemicycle ($180^{\circ}\leq\varphi_{\text{B}}<360^{\circ}$), the opposite effect occurs.
	
\begin{figure}[hbtp!]
\includegraphics[width = 8.5 cm]{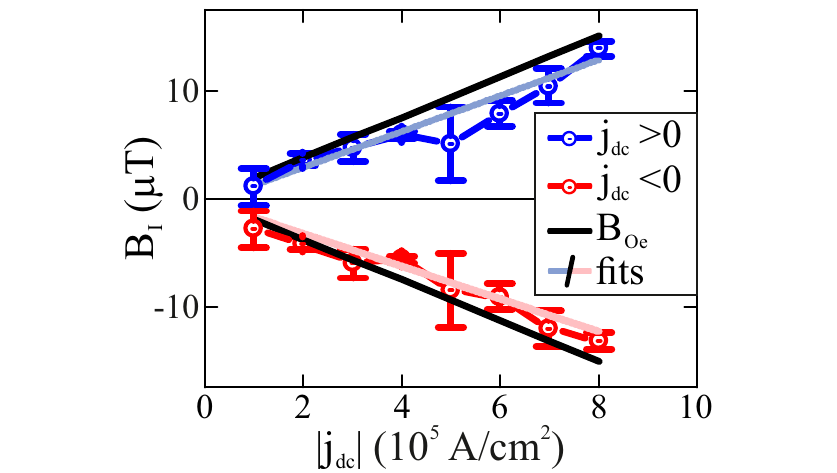}
\caption{\label{fig:current_linear} Current dependence of $B_{\rm I}$ in YIG(6~nm)/Pt(3~nm) for positive and negative dc offsets. The red and blue lines are linear fits to the data. The thin black lines show the Oersted field calculated from Amp\`{e}re's law.}
\end{figure}

%%%%%%%%%%%%%%%%%%%%%%%%%%%
In order to quantify $B_{\rm I}$, we performed a series of measurements for positive and negative dc offsets, shown in Fig.~\ref{fig:current_shift}~(e). We then fitted the $R_{\rm xy}$ curves using the energy functional from Eq.~(\ref{eq:E}) while keeping $B_{\rm K}$ and $\varphi_{\rm EA}$ equal to the values determined in the absence of a dc current and $B_{\rm I}$ as the only free parameter. The simulations, shown in Fig.~\ref{fig:current_shift}~(f), reproduce well the current-dependent switching observed in Fig.~\ref{fig:current_shift}~(e). Overall, the model supports the presence of a field $B_{\rm I} \parallel \pm y$ for a dc current $j_{\text{dc}} \parallel \pm x$, which has the same symmetry as the Oersted field expected from the current flowing in the Pt layer. The current dependence of $B_{\rm I}$, reported in Fig.~\ref{fig:current_linear}, further shows that $B_{\rm I}$ scales linearly as a function of $j_{\text{dc}}$ and that its amplitude is comparable with the Oersted field calculated from Amp\`{e}re's law as $B_{\rm Oe} = \mu_0 j_{\rm dc}t_{\rm Pt}/2 \approx 0.19$~mT for $j_{\rm dc}=10^7$~A/cm$^2$ (thin black line), where $t_{\rm Pt}$ is the thickness of Pt and $\mu_0$ denotes the vacuum permeability.
%A more extensive discussion of the origin of $B_{\rm I}$ will be given at the end of the paper.}

\begin{figure}[hbtp!]
\includegraphics[width=8.5cm]{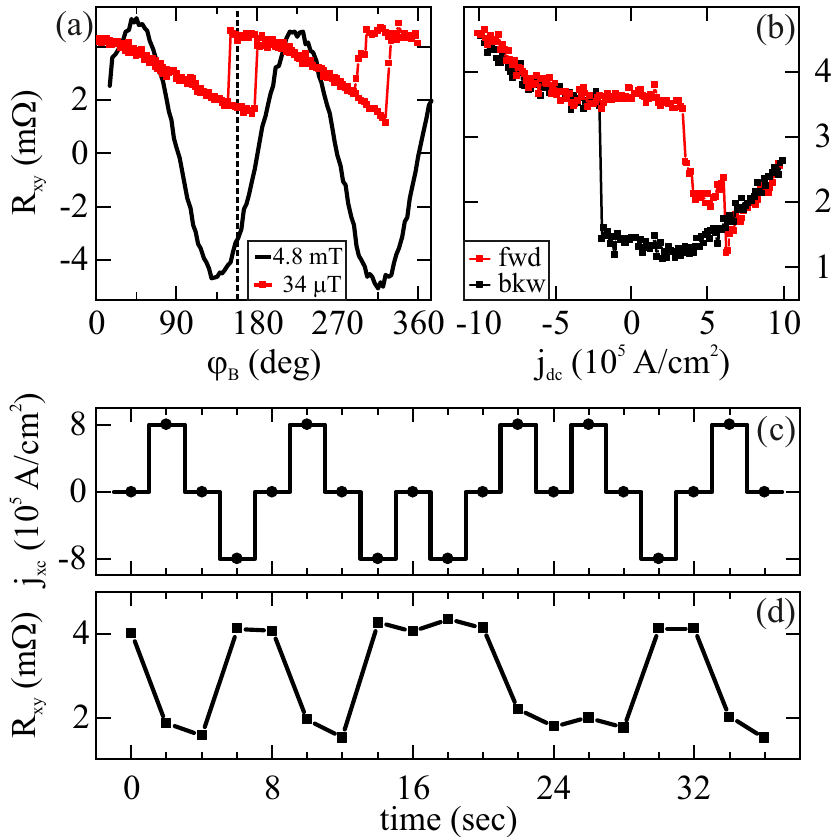}
\caption{\label{fig:hys_and_current_induced_switching_sep} (a) $R_{\rm xy}$ of YIG(7~nm)/Pt(3~nm) as a function of $\varphi_{\text{B}}$ at $B_{\text{ext}}= 4.8$~mT (black line) $B_{\text{ext}}= 34$~$\mu$T (red line). The dashed line indicates the value of $\varphi_{\text{B}}$ used for current-induced switching. (b) $R_{\rm xy}$ during a forward (red curve) and backward dc current sweep (black curve). (c,d) Current sequence and $R_{\rm xy}$ measured at $\varphi_{\text{B}} = 160^{\circ}$ and $B_{\text{ext}}= 34$~$\mu$T.}
\end{figure}

%%%%%%%%%%%%%%%%%%%%%%%%%%%% Switching by current %%%%%%%%%%%%%%%%%%%%%%%%%%
The presence of uniaxial in-plane anisotropy and the finite $B_{\rm I}$ allow us to switch the YIG magnetization by ramping the dc current in Pt. To enable the current-induced switching, we select a configuration in which the magnetization is bistable, namely the hysteretic region of $R_{\text{xy}}$ shown by the red curve in Fig.~\ref{fig:hys_and_current_induced_switching_sep} (a). We thus fix $B_{\rm ext}=34$~$\mu$T at $\varphi_{\text{B}}=160^{\circ}$ when sweeping from 360$^{\circ}$ to 160$^{\circ}$ which corresponds to the point indicated by the dashed line in Fig.~\ref{fig:hys_and_current_induced_switching_sep} (a). In this configuration, the magnetization is tilted towards the hard axis.  We then ramp the dc current towards positive values and simultaneously record $R_{\text{xy}}$  [red curve in Fig. \ref{fig:hys_and_current_induced_switching_sep} (b)]. From our former analysis, we expect that $B_{\rm I}$ induces a tilt $\Delta\varphi$ that will eventually lead to switching. Indeed, when reaching $j_{\text{dc}}=5\times10^5$~A/cm$^{2}$, we observe a step-like decrease of $R_{\text{xy}}$ indicating the reversal of the magnetization, followed by a parabolic-like increase of $R_{\text{xy}}$ at higher current, which we assign to a tilt of the magnetization in areas close to the Hall cross that have not switched. When sweeping the current back to zero, $R_{\rm xy}$ remains in the low resistance level (black curve). Moreover, the resistance switches back to the initial value at $j_{\text{dc}}=-2\times10^5$~A/cm$^{2}$. This behavior is similar to that reported for the current-induced switching of strained GaMnAs layers, with the difference that $B_{\rm I}$ in GaMnAs originates from spin-orbit coupling rather than by the Oersted field.\cite{Chernyshov2009}

Figures~\ref{fig:hys_and_current_induced_switching_sep}~(c) and (d) further show that the switching is reproducible for a sequence of positive and negative current pulses. In particular, the high and low levels of $R_{\rm xy}$ reproduce the full excursion of the $R_{\rm xy}$ signal at $B_{\rm ext}=34$~$\mu$T [red curve in Fig.~\ref{fig:hys_and_current_induced_switching_sep} (a)] and persist at zero dc current confirming the remanent character of the switching. Moreover, applying two consecutive pulses with the same current polarity does not lead to an additional increase or decrease of $R_{\rm xy}$, suggesting that the switching occurs between well-defined magnetization states, suggesting that the reversal process involves a majoritary and reproducible portion of the magnetic layer in the proximity of the Hall cross. Additional effects due to Joule heating are neglected, since the temperature increase derived from measurements of the resistivity during current injection is lower than 1~K.

\begin{figure}[hbtp!]
    \includegraphics[width = 8.5 cm]{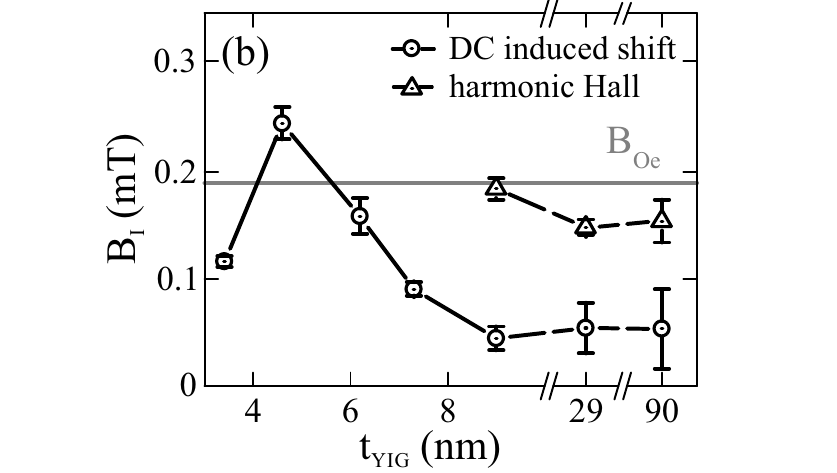}
    \caption{\label{fig:all_fields}Thickness dependence of the current-induced effective field $B_{\text{I}}$ measured by the angular shift method (dotted circles) and harmonic Hall voltage measurements (dotted triangles). The data are shown for a current density $j_{\text{dc}}=10^7$~A/cm$^2$.}
\end{figure}

%%%%%%%%%%%%%%%%%%%%%%%%%%%% Thickness dependence %%%%%%%%%%%%%%%%%%%%%%%%%%%
Before concluding, we discuss the origin of the current-induced field $B_{\rm I}$. As seen in Fig.~\ref{fig:current_linear}, $B_{\rm I}$ is only slightly smaller than $B_{\rm Oe}$, suggesting that $B_{\rm I}$ is dominated by the Oersted field, with possibly a small opposing spin-orbit effective field at the interface with Pt.\cite{Garello2013} This conclusion is consistent with earlier work on the current-induced ferromagnetic resonance of YIG/Pt bilayers.\cite{Fang2017} As the Oersted field acts on the entire magnetic volume of YIG, we also expect that $B_{\rm I}$ does not depend on the YIG thickness ($t_{\rm YIG}$). Measurements of the angular shifts $\Delta\varphi$ as a function of $t_{\rm YIG}$, however, give values of $B_{\rm I}$ that vary significantly between $t_{\rm YIG}=$3.5~nm and 7~nm, and finally saturate to about 0.05~mT/(10$^7$ Acm$^{-2}$) for $t_{\rm YIG}\geq 9$~nm (dotted circles in Fig.~\ref{fig:all_fields}), which is much smaller than $B_{\text{Oe}}\approx0.19$~mT/($10^{7}$ Acm$^{-2}$) (gray line in Fig.~\ref{fig:all_fields}). Whereas the increase of $B_{\rm I}$ between $t_{\rm YIG}=$3.5~nm and 4.5~nm can be attributed to a reduction of the interfacial spin-orbit effective field, which has opposite direction relative to $B_{\text{Oe}}$ and scales as 1/$t_{\rm YIG}M_s$, the monotonic decrease of $B_{\rm I}$ observed at $t_{\rm YIG}>4.5$~nm has apparently no explanation. Furthermore, harmonic Hall voltage measurements\cite{Garello2013,Avci2014b} of $B_{\rm I}$ performed on thick YIG samples yield values of $B_{\rm I}$ that are consistent with $B_{\text{Oe}}$ (dotted triangles in Fig.~\ref{fig:all_fields}), in clear contrast with $B_{\rm I}$ determined from the angular shifts of the hysteretic $R_{\rm xy}$ curves. This apparent discrepancy can be reconciled by taking into account the pinning of domain walls, which influences the magnetization reversal and hence the values of $B_{\rm I}$ determined using the angular shift method. Indeed, x-ray photoelectron emission microscopy shows that the domain morphology of YIG undergoes a transition around $t_{\rm YIG}= 9$~nm, changing from an irregular elongated pattern to 100~$\mu$m-wide pinned zigzag domains.\cite{Mendil2019} We therefore conclude that $B_{\rm I}$ originates mostly from the Oersted field, and that its effect on the magnetization is highly sensitive to the local pinning field, which depends strongly on $t_{\rm YIG}$. Finally, we note that the harmonic Hall voltage measurements were not feasible in the thinner samples due to additional effects overlapping with the Oersted field and spin-orbit torques, which are likely due to the small coercivity of the layers and prevent a reliable analysis of the data.

%%%%%%%%%%%%%%%%%%%%%% Conclusions %%%%%%%%%%%%%%%%%%%%%%%%%%%%%%%%%
In summary, we have shown that the current-induced effective field $B_{\rm I}$ is sufficient to reversibly manipulate the direction of the magnetization in YIG/Pt bilayers with in-plane anisotropy in the presence of a weak static external field. In YIG films thicker than 4~nm, $B_{\rm I}$ is consistent in sign and magnitude with the Oersted field generated by the current flowing in the Pt layer. Current-induced switching is achieved at an extremely small current density ($2\times10^5$~A/cm$^{2}$), which is two orders of magnitude smaller compared to the dc current switching of metallic ferromagnets such as Pt/Co\cite{Garello2014}, ferrimagnets such as Pt/GdCo\cite{Mishra2017}, and even thulium iron garnet/Pt.\cite{Avci2017} We attribute this difference to the extremely small uniaxial anisotropy and depinning field of YIG compared to ferro- and ferrimagnets with perpendicular anisotropy. The switching efficiency decreases in films thicker than 7~nm, which we attribute to a change of the domain morphology and increased pinning of the magnetic domain walls.~\cite{Mendil2019} Strain engineering of YIG thin films may be used to further tailor the magnetic anisotropy\cite{Wang2014h} and hence the switching behavior of YIG in response to current-induced fields of either Oersted or spin-orbit origin. Our results should also be taken as a cautionary warning about the possible undesired switching of YIG at current densities commonly used to excite and sense the magnetization of Pt/YIG bilayers.

\begin{acknowledgments}
We acknowledge financial support by the Swiss National Science Foundation under grant no. 200020-172775. We thank Can Onur Avci for valuable discussions.
\end{acknowledgments}

\end{document}